\documentclass[prc,twocolumn,showpacs,preprintnumbers,amsmath,amssymb,superscriptaddress,floatfix,nofootinbib]{revtex4}
\usepackage{graphicx}
\usepackage{epsfig}
\usepackage{amsmath}
\newcommand{\slashed}[1]{#1\!\!\!/}

\newcommand{\tabincell}[2]{\begin{tabular}{@{}#1@{}}#2\end{tabular}}

\def\prl#1{Phys.\ Rev.\ Lett.\ {\bf #1}}

\def\prc#1{Phys.\ Rev.\ C\ {\bf #1}}

\def\etal{{\em et al.}}

  \def\CL{{\cal L}}

\def\be{\begin{equation}}
\def\ee{\end{equation}}
\def\Be{\begin{eqnarray}}
\def\Ee{\end{eqnarray}}
\def\ba{\begin{array}}
\def\ea{\end{array}}

\begin{document}
\title{Nucleon resonance production in the $\gamma p \to p\eta \phi$ reaction}

\author{Jin-Quan Fan} \author{Shao-Fei Chen} \author{Bo-Chao Liu}
\email{liubc@xjtu.edu.cn}

\address{School of Science, Xi'an Jiaotong University, Xi'an, Shaanxi 710049, China}

\begin{abstract}
In this work, we perform a study of nucleon resonance production in
the $\gamma p \to p\eta\phi$ reaction within an effective Lagrangian
approach. In our model, we consider the excitation of the
$N^*(1535)$, $N^*(1650)$, $N^*(1710)$ and $N^*(1720)$ in the
intermediate state and the background term. We find that this
reaction is dominated by the excitation of the $N^*(1535)$ in the
near threshold region. Especially, we study the possible role of the
scalar meson exchange in this reaction. It is found that the
$f_0(980)$ exchange may give a significant contribution and the
parity asymmetry can be used to identify its role in this reaction.

\end{abstract}
\maketitle

\section{Introduction}
To study the properties of nucleon resonance is a central task in
hadronic physics. Despite decades of studies of the nucleon
resonances, there are still some controversies involving the nature
and structure of some nucleon resonances. One outstanding example is
the $N^*(1535)$. In addition to the conventional three quark
picture, there are evidences that in the $N^*(1535)$ there may be a
large mixture of the 3-quark and the molecular or penta-quark
components\cite{oset2002,Liu2006}. In the new picture, one can not
only naturally explain why the $N^*(1535)$ has a strong coupling to
$N\eta$ but also predict a large coupling of this resonance with
other strange channels\cite{Zou2009}. To verify these theoretical
models and their predictions, it is necessary to study this
resonance in various reactions.

In recent years, the photoproduction processes are widely employed
to investigate the properties of nucleon
resonance\cite{Burkert2018,Anisovich2017,Mattione2017,Kamano2013,Gothe2016}.
Up to now, most of these studies concentrate on the single meson
production process. However, studying nucleon resonance in some
multi-meson production processes is also interesting and can benefit
the understanding of the properties of nucleon resonances. For
example, in Ref.\cite{lv} it was shown that, due to the special
reaction mechanisms, the reaction $\gamma p\to \phi K^+\Lambda$ is
suitable for studying the $N^*(1535)K\Lambda$ coupling which is
difficult to be studied in the single meson production process. Such
example tells us, by inspecting the production mechanisms of nucleon
resonance in multi-meson production processes, it is possible to
learn about their couplings with various channels\cite{Liu2017,lv}.
Therefore, these studies are helpful for us to better understand the
nature and properties of the $N^*$s.

In this paper, we present the results for the study of the $\gamma p\to \eta \phi p$
 reaction using an effective Lagrangian approach. We
consider the contributions from the $N^*(1535)$, $N^*(1650)$,
$N^*(1710)$ and $N^*(1720)$ in the intermediate state, which then
decay into the $\eta N$ in the final state. The resonance
contribution in the $\phi\eta$ and $\phi p$ channels is ignored in
this work, because in the energy region under study no significant
resonance signals are found in these two channels\cite{PDG}. According to this
assumption, it seems that the present reaction may be a good place
to study the excitation mechanisms of the nucleon resonances in the
$\gamma p \to \phi N^*$ process since the decay process of $N^*\to
\eta N$ is relatively well known. Such studies are not only
important for understanding the reaction mechanism itself, but also
helpful for learning about the coupling of nucleon resonances with
the exchanged particles. In the $\gamma p \to \phi N^*$ process, due
to the conservation of C parity, vector meson exchange is forbidden
in this reaction. Since the involved nucleon resonances in this work
have relatively large decay branch ratios to $N\pi$ and $N\eta$
channels, it is natural to expect that the $\pi$ and $\eta$
exchanges play important roles for the excitation of the $N^*$s in
this reaction. While, there are still some other possible
contributions from such as scalar meson($\sigma$, $f_0(980)$ and
$a_0(980)$) and axial vector meson($a_1(1235)$) exchanges. These
contributions were usually ignored in previous studies due to their
relatively large mass or the ignorance of their couplings with
$N^*$s. Certainly, these assumptions should be verified by the
experiments. In this work, we hope to address the role of
$f_0(980)$, which is denoted by $f_0$ in the rest of the paper, for
the excitation of the $N^*(1535)$ in this reaction. To evaluate the
$f_0$ exchange contribution, the knowledge of the $\phi f_0\gamma$
and $N^*(1535)Nf_0$ couplings is essential. The $\phi f_0\gamma$
coupling can be extracted from the $\phi$ radiative decay to $\gamma
f_0$\cite{Kalashnikova2005}. While, the coupling of the $N^*(1535)$
with $Nf_0$ was rarely studied in previous works. However, if the
$N^*(1535)$ has a significant strange component and tends to have a
strong coupling with strange channel, it is possible that the
$N^*(1535)$ also has a large coupling with the $Nf_0$ channel since
the $f_0$ is believed to be a $K\bar K$ molecular state. If so, this
means that the present reaction may be a good place to look for the
evidence of the $N^*(1535)Nf_0$ coupling. As for the $a_0(980)$ and
$a_1(1235)$, we choose to ignore their contributions at present
partially because their couplings to the $N^*(1535)$ are totally
unknown and partially because their couplings with $\phi\gamma$ are
weak compared to the $f_0$\cite{PDG,Du2010}. The other scalar meson
$\sigma$ is also ignored due to the weak coupling of the
$\phi\gamma\sigma$ vertex\cite{PDG,Kucukarslan2013}. Next important
question is how to separate the scalar exchange contribution from
others. Inspired by some previous work\cite{Tannoudji,Schilling}, we
find spin observable is very useful for this purpose. As we know,
the spin density matrix elements(SDMEs) of $\phi$ meson can be
extracted from its decay angular distribution. By analyzing the
SDMEs of the $\phi$ meson, it is possible to learn the information
about the exchanged particles. Such method was already used in
Ref.\cite{Oh2006,Oh2012} to identify the role of the scalar $\kappa$
meson exchange. In this work, we will show that the SDMEs may also
offer useful information about the exchanged meson in the nucleon
resonance production processes.

The rest of this paper is organized as follows. In Sec.II, we
present the theoretical formalism used in the calculations.
Numerical results and discussions are presented in Sec. III,
followed by a summary in the last section.
\begin{figure*}[htbp]
\centering
\includegraphics[scale=0.12]{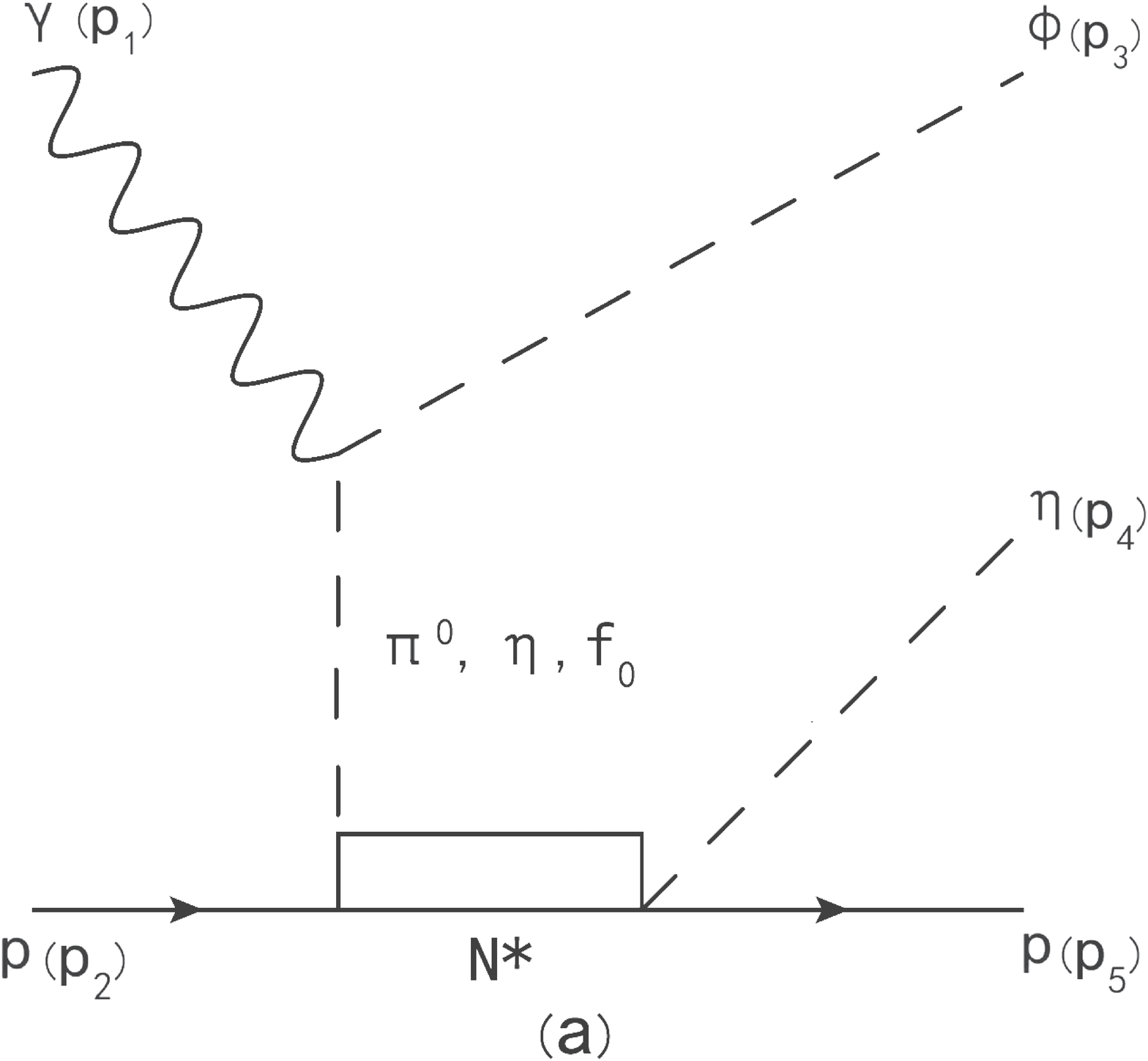}
\includegraphics[scale=0.33]{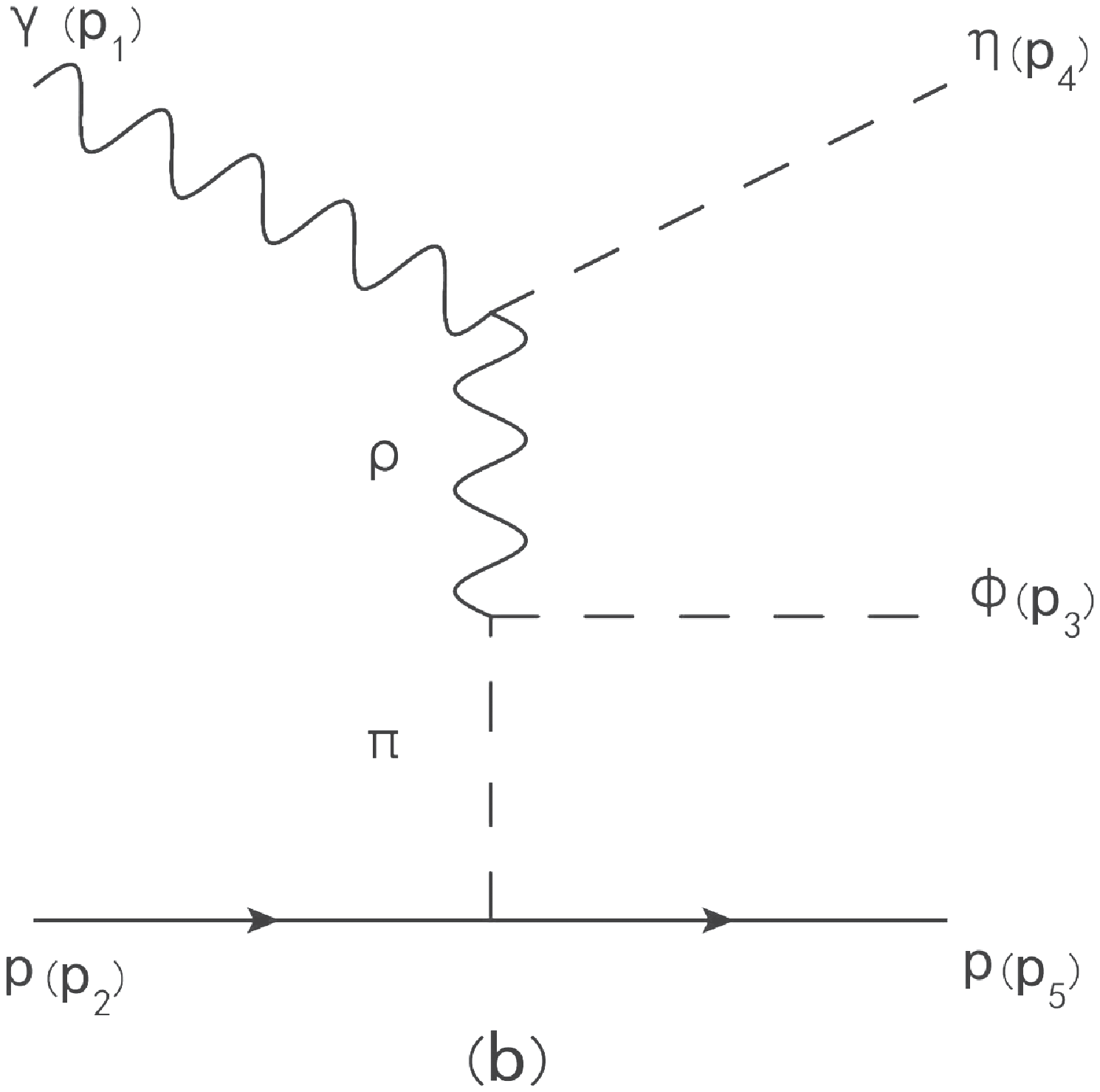}
\caption{Feynman diagrams for the ${\gamma}p \to {\phi}{\eta}p$
reaction.} \label{feynfig}
\end{figure*}

\section{model}
In our model, the Feynman diagrams for the $\gamma p \to\phi \eta p$
reaction can be depicted by the Fig.\ref{feynfig}.  As mentioned in
the introductory part, here we only consider the $\pi$, $\eta$ and
$f_0$ exchanges for the excitation of nucleon resonances since other
meson exchanges are either forbidden or expected to be unimportant.

To calculate the Feynman diagrams shown in Fig.\ref{feynfig}, we
need the Lagrangian densities for the various vertices. The
following Lagrangian densities are employed in this
work\cite{lv,Lee2017,Zou2003}: \Be
\mathcal{L}_{{\pi}NN} &=& - \frac{{{g_{\pi NN}}}}{{2{m_N}}}\bar N {\gamma _5}{\gamma _\mu }{\partial ^\mu }{\pi}N,\\
\CL_{\gamma {\pi}\phi } &=& {e \over {{m_\phi }}}{g_{\phi \gamma \pi
}}{\varepsilon ^{\mu \nu \alpha \beta }}{\partial _\mu }{\phi
_\nu }{\partial _\alpha }{A_\beta }{\pi},\\
\CL_{\gamma \eta \phi} &=& {e \over {{m_\phi }}}{g_{\phi \gamma \eta
}}{\varepsilon ^{\mu \nu \alpha \beta }}{\partial _\mu }{\phi _\nu
}{\partial _\alpha }{A_\beta }\eta,\\
\CL_{\rho_0 \eta \gamma} &=& \frac{e}{{{m_\rho }}}{g_{\rho \eta
\gamma}}{\varepsilon ^{\mu \nu \alpha \beta }}{\partial _\mu
}{\rho_\nu }{\partial _\alpha }A_\beta \eta,\\
\CL_{\phi \rho_0 \pi} &=& \frac{e}{{{m_\phi }}}{g_{\phi \rho
\pi}}{\varepsilon ^{\mu \nu \alpha \beta }}{\partial _\mu
}{\phi_\nu }{\partial _\alpha }\rho_\beta \pi,\\
\CL_{\phi f_0 \gamma}&=&\frac{e}{m_\phi}g_{\phi f_0 \gamma}{\partial
^\alpha}{\phi ^\beta}({\partial _\alpha}A_\beta-{\partial
_\beta}A_\alpha)f_0,\\
\CL_{f_0 N N^*_{1535}}&=& g_{f_0 N N^*_{1535}}f_0 \bar{N}
\gamma_5 N^* + h.c. ,\\
\CL_{MN{N_{1535}^*}} &=& i{g_{MN{N_{1535}^*}}}{\bar N^*}MN + h.c.,\\
\CL_{MN{N_{1650}^*}} &=& i{g_{MN{N_{1650}^*}}}{\bar N^*}MN + h.c.,\\
\CL_{MNN_{1710}^*} &=&  - \frac{g_{MNN_{1710}^*}}{{m_N} +
{m_{N^*}}}{\bar{N}^*}{\gamma _5}{\gamma _\mu }{\partial ^\mu }MN
+ h.c.,\\
\CL_{MNN_{1720}^*} &=&  -
\frac{{{g_{MNN_{1720}^*}}}}{{{m_M}}}{\bar{N}}_\mu ^*{\partial ^\mu
}MN + h.c. ,\Ee where $e = \sqrt {4\pi/137 } $, $\phi_\mu$ is the
$\phi$ meson field, M stands for $\pi$ or $\eta$ field and $A_\mu$
is the photon field. The coupling constant $g_{\pi NN}$ is taken
from Ref.\cite{Xie2008,Xie2013} with $g_{\pi NN}=13.45$. Other
coupling constants can be determined
through the following formulae, \Be
\Gamma [V \to P \gamma] &=& \frac{{{e^2}g_{V P \gamma}^2}}{12\pi
}\frac{{\left| {\vec{p}}
\right|}^3}{m_\rho ^2},\\
\Gamma [\phi  \to \rho_0 \pi] &=& \frac{{{e^2}g_{\phi \rho
\pi}^2}}{12\pi }\frac{{\left| {\vec{p}}
\right|}^3}{m_\phi ^2},\\
\Gamma [\phi \to f_0 \gamma]&=&\frac{{e^2}g_{\phi f_0
\gamma}^2}{{12\pi }}\frac{{{{\left| {{\vec{p}}}
\right|}^3}}}{{m_\phi ^2}}\label{f0phigamma},\\
\Gamma [{N_{\frac{1}{2}^-}^*} \to P N] &=& \frac{\kappa
{g_{PN{N_{1/2^-}^*}}^2}}{{4\pi }}\frac{{({E_N} +
{m_N})}}{{{m_{{N^*}}}}}\left| {{\vec{p}}} \right|,\\
\Gamma [{N_{1710}^*} \to P N] &=& \frac{\kappa
{g_{PN{N_{1710}^*}}^2}}{{4\pi }}\frac{({E_N} -
{m_N})}{m_{N^*}}\left| {{\vec{p}}} \right|,\\
\Gamma [{N_{1720}^*} \to P N] &=& \frac{\kappa
{g_{PN{N_{1720}^*}}^2}}{{12\pi }}\frac{{({E_N} +
{m_N})}}{{m_{N^*}}m_{P}^2}{\left| {{\vec{p}}} \right|}^3, \Ee where
p denotes the magnitude of the momentum of final particles  in the
center of mass frame. P and V denote the pseudoscalar and vector
meson respectively. $\kappa$ is a constant which equals 1 for the
$\eta$ exchange and 3 for the $\pi$ exchange. The obtained coupling
constants are listed in Tab.\ref{cc}.

\begin{table}[htbp]
\centering \caption{Coupling constants used in this work. The
experimental decay branch ratios are taken from Ref.\cite{PDG}.}
\label{cc}
\begin{tabular}{ccccc}
\hline  
\hline  
State&\tabincell{c}{Width\\(Mev)}&\tabincell{c}{Decay\\channel}&\tabincell{c}{Adopt\\branching
ratio}&$g^2/4\pi$\\
\hline  
$\rho_0$&$147.8$&$\eta \gamma$&$3.0 \times 10^{-4}$&$0.12$\\
$\phi$&$4.25$&$\pi \gamma$&$1.3 \times 10^{-3}$&$1.60 \times 10^{-3}$\\
&&$\eta \gamma$&$1.3 \times 10^{-2}$&$3.97 \times 10^{-2}$\\
&&$\rho \pi$&$0.15$&$3.55$\\
&&$f_0 \gamma$&$3.2 \times 10^{-4}$&$1.73$\\
$N^*(1535)$&$150$&$N \pi$&$0.42$&$3.43 \times 10^{-2}$\\
&&$N \eta$&$0.42$&$0.28$\\
$N^*(1650)$&$125$&$N \pi$&$0.60$&$3.73 \times 10^{-2}$\\
&&$N \eta$&$0.25$&$7.44 \times 10^{-2}$\\
$N^*(1710)$&$140$&$N \pi$&$0.13$&$0.10$\\
&&$N \eta$&$0.30$&$2.03$\\
$N^*(1720)$&$250$&$N \pi$&$0.11$&$2.04 \times 10^{-3}$\\
&&$N \eta$&$0.03$&$8.25 \times 10^{-2}$\\
\hline  
\hline  
\end{tabular}

\end{table}

As we are not dealing with point-like particle here, the form
factors are necessary to be introduced. In the present work, we
choose the following form factor for the baryon exchange diagrams as
in Ref.\cite{Shklyar2005,Xie2012}

\begin{equation}
{F_B}(q_{ex},{m_{ex}}) = \frac{{\Lambda _B^4}}{{\Lambda _B^4 +
{{(q_{ex}^2 - m_{ex}^2)}^2}}}.
\end{equation}

For $\pi$ meson and $\eta$ meson exchange diagrams, we
adopt\cite{Oh2008}

\begin{equation}
{F_M}(q_{ex},{m_{ex}}) = {(\frac{{\Lambda _M^2 -
m_{ex}^2}}{{\Lambda _M^2 - q_{ex}^2}})^2}.
\end{equation}
While, we use
\begin{equation}
{F_V}(q_{ex}) = {(\frac{{\Lambda _V^2}}{{\Lambda _V^2 -
q_{ex}^2}})^2}
\end{equation}
for $\rho$ meson exchange\cite{Oh2011}. The $q_{ex}$ and $m_{ex}$
are the four-momentum and mass of the exchanged particle,
respectively. As for the cutoff parameters, we take ${\Lambda
_{\pi}} = {\Lambda _{\eta}} = 1.3$ GeV and ${\Lambda _{\rho}}=1.2$
GeV for meson exchanges\cite{Liu2017,Xie2008} and ${\Lambda
_{B}}=2.0$ GeV \cite{lv}for baryon exchanges.

The propagators for the exchanged particles are used as
\begin{equation}
{G_0}(q) = \frac{i}{{{q^2} - m^2}}
\end{equation}
\\for $\pi$ and $\eta$,
\begin{equation}
{G_1^{\mu\nu}}(q) = -\frac{i(g^{\mu \nu}- {q^\mu
q^\nu}/{q^2})}{{q^2} - m^2}
\end{equation}
\\for $\rho$,
\begin{equation}
G_{\frac{1}{2}}(q) = \frac{i(\slashed{q} + m)}{{q^2} - {m^2} +
im\Gamma }
\end{equation}
\\for the spin-1/2 baryon and
\begin{equation}
G_{\frac{3}{2}}^{\mu \nu }(q) = \frac{{i(\slashed{q} + m){P^{\mu \nu
}}(q)}}{{{q^2} - {m^2} + im\Gamma }}
\end{equation}
for the spin-3/2 baryon with
\begin{equation}
{P^{\mu \nu }}(q) =  - {g^{\mu \nu }} + \frac{1}{3}{\gamma ^\mu
}{\gamma ^\nu } + \frac{1}{{3m}}({\gamma ^\mu }{q^\nu } - {\gamma
^\nu }{q^\mu }) + \frac{2}{{3{m^2}}}{q^\mu }{q^\nu }.
\end{equation} Here q, m, and $\Gamma$ are the four-momentum, mass and
width of the exchanged particle.

By using the above effective Lagrangian densities and the
propagators, we can get the scattering amplitude of the Feynman
diagrams shown in Fig.\ref{feynfig}. The corresponding amplitudes
for Fig.\ref{feynfig}a are
\begin{equation}
\begin{aligned}
M_{N^*_{\frac{1}{2}^-}} =& \frac{e{g_{\phi \gamma P
}}{g_{N{N^*}P}}{g_{\eta{N^*}N}}}{m_\phi[{{{({p_3} - {p_1})}^2} -
m_P^2}]}{\bar{u}}({p_5},{s_5})G_{\frac{1}{2}}(q_{N^*})\\
&\times  u({p_2},{s_2}) {\varepsilon ^{\mu \nu \alpha \beta
}}{P_{3\mu }}\varepsilon _\nu ^*({p_3},{s_3}){p_{1\alpha
}}\varepsilon _\beta ({p_1},{s_1})\\
&\times {F_B}(q_{N^*},{m_{N^*}}){F_M}(q_{P},{m_{P}}),
\end{aligned}
\end{equation}

\begin{equation}
\begin{aligned}
M_{N^*_{\frac{1}{2}^+}} =& \frac{{e{g_{\phi \gamma P }}{g_{N{N^*}P
}}{g_{\eta
{N^*}N}}}}{{{m_\phi}{m_p+m_{N^*}}^2}}{\bar{u}}({p_5},{s_5}){\gamma
_5} {\slashed{p}_4}G_{\frac{1}{2}}(q_{N^*})\\
\times&{\gamma_5}({\slashed{p}_3} - {\slashed{p}_1})u({p_2},{s_2})
\frac{{F_B}(q_{N^*},{m_{N^*}}){F_M}(q_{P},{m_{P}})}{{{({p_3} - {p_1})}^2} - m_P ^2}\\
&\times{\varepsilon ^{\mu \nu \alpha \beta }}{P_{3\mu }}\varepsilon
_\nu ^*({p_3},{s_3}){p_{1\alpha }}\varepsilon _\beta ({p_1},{s_1}),
\end{aligned}
\end{equation}

\begin{equation}
\begin{aligned}
{M_{{N^*_{\frac{3}{2}^+}}}} =& \frac{{e{g_{\phi \gamma
P}}{g_{N{N^*}P}}{g_{\eta {N^*}N}}}}{{{m_\phi }{m_\eta
}{m_P}}}{\bar{u}}({p_5},{s_5}){p_{4\mu
}}G_{\frac{3}{2}}^{\mu \nu }(q_{N^*})\\
\times &{({p_1} - {p_3})_\nu }u({p_2},{s_2})
\frac{{F_B}(q_{N^*},{m_{N^*}}){F_M}(q_{P},{m_{P}})}{{{{({p_3} - {p_1})}^2} - m_P^2}}\\
&\times{\varepsilon ^{\mu \nu \alpha \beta }}{p_{3\mu }}\varepsilon
_\nu ^*({p_3},{s_3}){p_{1\alpha }}\varepsilon _\beta ({p_1},{s_1}),
\end{aligned}
\end{equation}
\begin{equation}
\begin{aligned}
M_{f_0} =& \frac{-ie{g_{\phi \gamma f_0
}}{g_{N{N^*}f_0}}{g_{\eta{N^*}N}}}{m_\phi[{{{({p_3} - {p_1})}^2} -
m_{f_0}^2}]}{\bar{u}}({p_5},{s_5})G_{\frac{1}{2}}(p_4+p_5)\\
&\times  \gamma_5 u({p_2},{s_2}) ((p_3 \cdot p_1)(\varepsilon_3^* \cdot \varepsilon_1)-(p_3 \cdot \varepsilon_1)(p_1 \cdot \varepsilon_3^*))\\
&\times {F_B}(q_{N^*},{m_{N^*}}){F_M}(q_{f_0},{m_{f_0}}),
\end{aligned}
\end{equation}
where P denotes the exchanged $\pi$ or $\eta$ and the momentum of
individual particles is denoted as in Fig.\ref{feynfig}. The
corresponding amplitude for Fig.\ref{feynfig}b can be written as
\begin{equation}
\begin{aligned}
M_t =& -\frac{{{ie^2}{g_{\phi \rho \pi }}{g_{\eta \rho \gamma
}}{g_{\pi NN}}}}{{2{m_p}{m_\phi }{m_\rho
}}}{\bar{u}}({p_5},{s_5}){\gamma _5}({\slashed{p}_2} -
{\slashed{p}_5})u({p_2},{s_2})\\
&\times \frac{{F_V}(q_{\rho}){F_M}(q_{P}^2,{m_{P}})}{{{{({p_5} -
{p_2})}^2} - m_\pi ^2}}{\varepsilon ^{\mu \nu \alpha \beta
}}{P_{3\mu }}\varepsilon _\nu
^*({p_3},{s_3})\\
&\times {({p_1} - {p_4})_\alpha }{G_1^{\beta \delta}}(p_4-p_1)
{\varepsilon ^{\rho\sigma\gamma\delta}}{p_{1\rho}}{\varepsilon
_\sigma}({p_1},{s_1}){({p_1} - {p_4})_\gamma}.
\end{aligned}
\end{equation}

\section{results and discussion}

\begin{figure}[htbp]
\small
\centering
\includegraphics[scale=0.3]{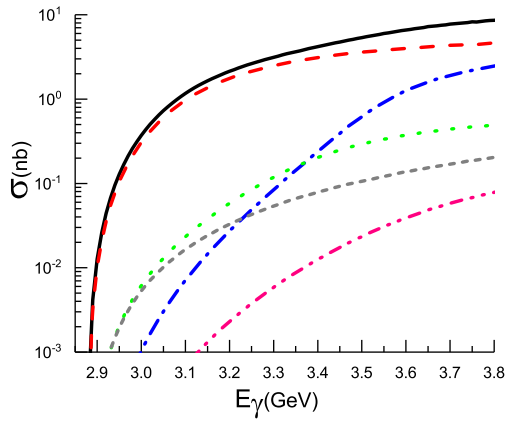}
\caption{Total cross section vs the beam energy $E_\gamma$ for the
${\gamma}p \to {\phi}{\eta}p$ reaction without considering the $f_0$
exchange. The red-dashed, green-dotted, blue-dash-dotted,
pink-dash-dot-dotted lines represent the contribution of the
$N^*(1535)$, $N^*(1650)$, $N^*(1710)$ and $N^*(1720)$, respectively.
The gray-short-dashed line stands for the t channel background
contribution. Their sum is shown by the solid black line.}
\label{xsection}
\end{figure}

\begin{figure}[htbp]
\centering
\includegraphics[scale=0.3]{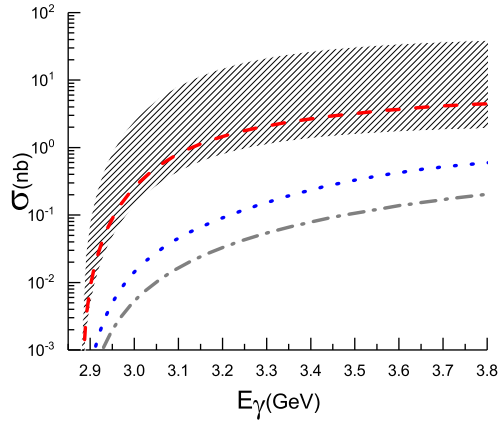}
\caption{Individual contributions of various meson exchanges. The
red-dashed, blue-dotted and gray-dash-dotted lines indicate the
contributions of the $\eta$, $\pi$ exchanges and background term,
respectively. The shadow area represents the $f_0$ exchange
contribution with $\Lambda_{f_0}$ varying from 1.3 GeV to 1.8 GeV.
}\label{cutoff}
\end{figure}

With the formulae given in the last section, the total and
differential cross sections can be calculated. Firstly, we consider
the case that the $f_0$ exchange contribution is not considered. In
Fig.\ref{xsection}, we show the contributions of the various nucleon
resonances and the background contribution. It is obvious that the
$N^*(1535)$ plays a dominant role in this reaction near threshold.
 The dominant role of the $N^*(1535)$ is mainly
attributed to its large coupling with the $N\eta$ channel. Other
nucleon resonances and the background term only play minor roles in
this reaction. While, with increasing energy, the $N^*(1710)$ will
become important gradually. The interference terms among the
individual amplitudes are usually important for understanding the
experimental data. Most of the coupling constants used in our work
are extracted from the decay width, in which way the relative phase
of the couplings can not be determined. Fortunately, due to the
dominant role of the $N^*(1535)$ in the energy region under study,
we do not expect the interference effects are very significant here.
In fact, we have tried to alter the phase by multiplying a factor of
-1 and no significant changes in the results have been found. It
should also be noted that, to evaluate the $\phi\rho\pi$ coupling
constant, we adopt the decay branch ratio of $\phi\to\rho\pi$ at the
upper limit suggested by the PDG\cite{PDG} which means the
background contribution could be even smaller.

Now we come to discuss the role of the $f_0$ exchange in this
reaction. To evaluate the contribution from the $f_0$ exchange, it
is necessary to identify the coupling constant of the
$N^*(1535)Nf_0$ vertex and the cutoff parameter in the form factor.
Up to now, these two parameters are still rarely known. Here we make
the guess that $g_{N^*(1535)Nf_0}$ has a similar magnitude as the
$g_{N^*(1535)N\sigma}$. For the latter, its value can be evaluated
through the partial decay width $\Gamma(N^*(1535)\to N \sigma)$ and
one can get $|g_{N^*(1535)N\sigma}|=2.55$\cite{Lee2017}. As a rough
estimate of the $f_0$ exchange contribution, we think this
substitute is plausible since the $N^*(1535)$ is expected to have an
even stronger coupling with final state containing significant
strange component such as $Nf_0$. On the other hand, for the cutoff
parameter $\Lambda_{f_0}$, we study the dependence of the results on
this parameter explicitly by varying it from 1.3 GeV to 1.8 GeV.

In Fig.\ref{cutoff}, we show the individual contributions of various
meson exchanges and the $\Lambda_{f_0}$ dependence of the $f_0$
exchange contribution. It is found that the role of the $\eta$
exchange is much more important than that of the $\pi$ exchange in
this reaction as expected. While, the role of $f_0$ exchange is
dependent on the adopted value of the $\Lambda_{f_0}$. However, even
with a relatively small value of $\Lambda_{f_0}$, i.e.
$\Lambda_{f_0}=1.3$ GeV, the $f_0$ exchange contribution is still
comparable to the $\eta$ exchange contribution and larger than the
contribution of the $\pi$ exchange and the background term. The
significant $f_0$ exchange contribution can be attributed to the
relatively large $\phi f_0\gamma$ coupling and the p-wave nature of
the $N^*(1535)Nf_0$ coupling. Compared to the s-wave coupling of the
$N^*(1535)N\eta$ vertex, the p-wave $N^*(1535)Nf_0$ coupling is
amplified due to the large threshold momentum of this reaction,
which enhances the $f_0$ exchange contribution. Note that to extract
the $\phi f_0\gamma$ coupling constant we use Eq.\ref{f0phigamma}
and the decay branch ratio of $\phi\to \gamma f_0$, where the
adopted value of the $f_0$ mass is essential in the calculation
since $\phi$ lies near the $f_0\gamma$ threshold. In the above
calculations, we take the mass of $f_0$ as 0.99
GeV\cite{PDG,Lee2017}. If a smaller mass, such as $M_{f_0}=0.98$
GeV, is adopted, the $g_{\phi f_0\gamma}^2$ will become a factor of
2-3 smaller than the value used here.
\begin{figure}[htbp]
\centering
\includegraphics[scale=0.3]{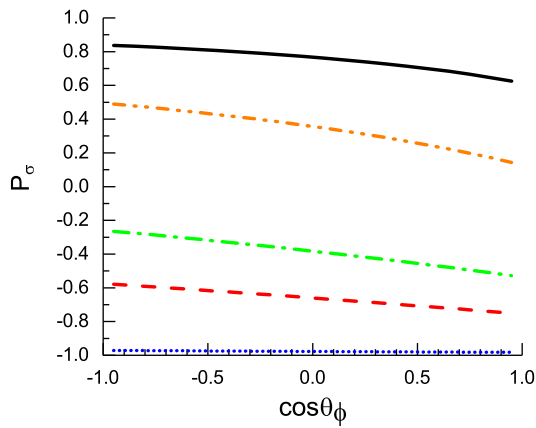}
\caption{Predictions for the $P_\sigma$ at $P_{\gamma}=3.0$ GeV. The
black-solid, orange-dash-dot-dotted and green-dash-dotted lines
stand for the results with $\Lambda_{f_0}=1.8$, 1.5 and $1.3$ GeV
respectively. The red-dashed line corresponds to the results using
$\Lambda_{f_0}=1.3$ GeV and a smaller mass of the $f_0$, i.e.
$m_{f_0}=0.98$ GeV, in the calculation. And the blue-dotted line
indicates the result without considering the $f_0$ exchange
contribution.}\label{psigma}
\end{figure}

Based on the results shown above, it seems that with the current
parameters the $f_0$ exchange contribution is significant in the
present reaction. To identify the role of $f_0$ exchange in this
reaction, it is important to find some observable to separate the
various meson exchange contributions. Inspired by the pioneer
works\cite{Schilling,Tannoudji}, it is found that the parity
asymmetry is suitable for this purpose. In Ref.\cite{Oh2006}, it has
been shown that this observable can be used to identify the scalar
exchange contribution. The parity asymmetry is defined by the SDMEs
as
\begin{equation}
P_\sigma=2\rho_{1 -1}^{1}-\rho_{0 0}^{1},
\end{equation}
In the $\gamma$ N $\to$ V N reaction, it can be proven that for the
natural and unnatural exchanges the parity asymmetry $P_\sigma$
equals to 1 and -1 respectively. Since the scalar meson $f_0$ has
the natural parity and the pseudoscalar meson $\eta$($\pi$) has
unnatural parity, it is possible to distinguish their contributions
by measuring this observable. It should be pointed out that an
important difference between our work and previous ones is here we
deal with nucleon resonance in the intermediate state. However, this
difference is irrelevant to the purpose of the present work, since
the $P_{\sigma}$ is solely determined by the $\phi$-$\gamma$-Meson
vertex if only scalar or pseudoscalar meson exchanges are concerned.
Such argument is supported by the numerical results shown in
Fig.\ref{psigma}. In this figure, we present the predictions for the
$P_\sigma$ with $\Lambda_{f_0}=1.3$, $1.5$ and $1.8$ GeV. When we
take the $\Lambda_{f_0}=1.8$ GeV, the $f_0$ exchange dominates this
reaction and the $P_\sigma$ is about 0.8. If we take
$\Lambda_{f_0}=1.3$ GeV, the $f_0$ exchange contribution is smaller
than that of the $\eta$ exchange and the $P_\sigma$ is about $-0.4$.
When the $f_0$ exchange contribution is not considered at all, the
$P_\sigma$ approaches -1 in accordance with the expectations. Note
that in this figure we also present the result using a smaller
value, i.e. 0.98 GeV, for the mass of $f_0$ to check the dependence
of the results on the mass of $f_0$. As shown in the figure, it
seems even in this case the predictions for the $P_\sigma$ is still
distinct from the results without considering the $f_0$ exchange
contribution. Therefore, we conclude that the parity asymmetry
$P_\sigma$ is suitable to identify the scalar exchange contribution
in this reaction.

\section{summary}
In this work, we investigate the $\gamma p\to p \eta\phi$ reaction
within an effective Lagrangian approach. We consider the
contribution of the $N^*(1535)$, $N^*(1650)$, $N^*(1710)$ and
$N^*(1720)$ in the intermediate state and the background
contribution. It is found that the production of the $N^*(1535)$
dominates this reaction in the near threshold region. Especially, we
study the possible role of the scalar exchange for the excitation of
$N^*(1535)$ and find that the $f_0$ may play an important role here.
We also find the parity asymmetry $P_\sigma$ is sensitive to the
scalar exchange contribution and can be used to identify the role of
the scalar exchange in this reaction.

\begin{acknowledgements}
We acknowledge the supports from the National Natural Science
Foundation of China under Grants No. 11375137 and U1832160, the
Natural Science Foundation of Shaanxi Province under Grants No.
2015JQ1003 and the Fundamental Research Funds for the Central
Universities.
\end{acknowledgements}

\end{document}